\documentclass[aps,superscriptaddress,pra,twocolumn,footinbib]{revtex4-1}
\usepackage{amsmath,amssymb}
\usepackage{bm}
\usepackage{epsfig}
\usepackage{natbib}
\usepackage{hyperref}

\usepackage{graphicx}
\usepackage{epstopdf}
\usepackage{color,xcolor}
\usepackage{bm}
 \hypersetup{pdfstartview={FitH},pdfpagemode={UseNone},
            colorlinks,linkcolor=red, citecolor=blue, urlcolor=blue,
            bookmarks=true, bookmarksopen=true, pdfnewwindow=true}
             
\usepackage{verbatim} 
\usepackage{morefloats}
\usepackage{bibentry}
\usepackage[mathscr]{euscript}

\newcommand{\eps}{\varepsilon}

\renewcommand{\phi}{\varphi}

\begin{document}

\title{Multipolar second-harmonic generation from high-$Q$ quasi-BIC states in nanoresonators}

\author{\firstname{Irina I.} \surname{Volkovskaya}}
\affiliation{Institute of Applied Physics, Russian Academy of Science, Nizhny Novgorod 603950, Russia}
\author{Lei Xu}
\affiliation{School of Engineering and Information Technology, University of New South Wales, Canberra ACT 2600, Australia}
\author{Lujun Huang}
\affiliation{School of Engineering and Information Technology, University of New South Wales, Canberra ACT 2600, Australia}
\author{\firstname{Alexander I.} \surname{Smirnov}}
\affiliation{Institute of Applied Physics, Russian Academy of Science, Nizhny Novgorod 603950, Russia}
\author{Andrey Miroshnichenko}
\affiliation{School of Engineering and Information Technology, University of New South Wales, Canberra ACT 2600, Australia}
\email{andrey.miroshnichenko@unsw.edu.au}

\begin{abstract}

Multimode interference and multipolar interplay 
govern functionalities of optical 
nanoresonators and nonlinear nanoantennas.
Recently, it has been unveiled that high-quality supercavity modes in individual subwavelength dielectric particles can be
associated with the concept of bound states in the continuum (BIC). Due to 
their high finesse, excitation of such quasi-BIC states in isolated particles has been predicted to boost the
nonlinear frequency conversion at the nanoscale. 
Here, we  put  forward the multipolar model which captures the physics behind linear and nonlinear response driven by 
the high-$Q$ modes in 
nanoresonators. 
We show   
that formation of the quasi-BIC state in the AlGaAs nanodisk can  be  naturally  understood  through multipolar transformations of coupled leaky modes. 
In particular, the 
hybridized axially symmetric TE-polarized modes 
can be viewed as superpositions of multipoles, with a basis of parent multipoles constituted mainly by magnetic dipoles and octupole. 
The quasi-BIC point in the parameter space appears with increasing the order of the dominating multipole, where dipolar losses are totally suppressed.
The efficient optical coupling to this state is implemented via azimuthally polarized beam illumination  matching its  multipolar origin.
We establish a one-to-one correspondence between the standard phenomenological  non-Hermitian coupled-mode theory and multipolar models that enables transparent interpretation of scattering characteristics. Using our 
approach, we 
derive the multipolar composition of the generated
second-harmonic radiation from the AlGaAs nanodisk possessing bulk quadratic nonlinearity and validate it with full-wave numerical simulations. Back-action of the second-harmonic radiation onto the fundamental frequency is taken into account in the coupled nonlinear model with pump depletion. 
A hybrid metal-dielectric nanoantenna is proposed to augment the conversion efficiency up to tens of per cent due to 
increasing quality factors of the resonant states at the fundamental and the second-harmonic frequencies. 
With such nanoantennas, the reverse process of the decay of 
the pump field into subharmonics, that is nonlinear parametric downconversion, 
can be actualized.
Our findings delineate 
the in-depth conceptual framework 
and novel promising strategies in the 
design of functional elements for nonlinear nanophotonics applications.
\end{abstract}

\maketitle
\section{Introduction}
%%%%%%%%%%%%%%%%%%%%%
Controlling light at the nanoscale has been a vibrant field of research for many years motivated by its various applications for 
optical nanoantennas, 
integrated photonic circuitry, optical computing and high-speed ultrathin photonic devices~\cite{Novotny2011Antennas,novotny2012principles,menon2010nonlinear,smirnova2016multipolar,kuznetsov2016optically}. High-index dielectric nanostructures have emerged as a promising platform to enhance light-matter interactions at the nanoscale 
based on optically induced electric and magnetic Mie-type resonances~\cite{smirnova2016multipolar,kuznetsov2016optically}.
 Due to the strong near-field confinement and tailorable field distributions in the subwavelength regime, optically resonant dielectric nanostructures has offered 
 powerful tools to facilitate various nonlinear processes including nonlinear frequency conversion, wave mixing and ultrafast all-optical switching~\cite{shcherbakov2015ultrafast,smirnova2016multipolar,Shcherbakov2017Ultrafast,li2017nonlinear,keren2018shaping,rahmani2018nonlinear,sain2019nonlinear}. 
Resonant mechanisms of light localization in dielectric nanostructures, such as magnetic dipole resonance~\cite{shcherbakov2014enhanced,Carletti2015benhanced,xu2018highly}, nonradiating anapole state~\cite{miroshnichenko2015nonradiating,xu2018boosting,Baryshnikova2019}, magnetic Fano resonance~\cite{yang2015nonlinear,shorokhov2016multifold}, and topologically protected edge states~\cite{kruk2019nonlinear,smirnova2019third}, have been widely utilized for applications of nonlinear nanophotonics, such as nanoscale light sources~\cite{marino2019spontaneous}, imaging~\cite{rodrigues2014nonlinear}, sensing and advanced optoelectronic devices~\cite{menon2010nonlinear}. 
%%%%%%%%%%%%%%%%%%%%%%%%%%%%%%%%%%%%%%%%

Methods typically employed to describe nonlinear harmonic generation at the nanoscale are based on multipolar decomposition of the fields in spherical multipoles~\cite{smirnova2016multipolar,Smirnova2016acsphotonics,Smirnova2018,Frizyuk2019,Kruk2017}. This approach provides a transparent interpretation for the measurable far-field characteristics, such as the conversion efficiency and radiation patterns, since they are essentially determined by the interference of dominating multipolar modes~\cite{smirnova2016multipolar}. 
The nonlinear response depends on both incident polarisation and symmetry of the a specific material or composite structures, such as nanoparticle oligomers~\cite{CamachoMorales2016,Kruk2017,Kroychuk2019}.
For instance, the nonlinear response of disk-shaped nanoantennas made of noncentrosymmetric III-V semiconductors (GaAs or AlGaAs) grown along (100), (110) and (111) crystallographic directions exhibits four-fold, two-fold and continuous rotational symmetries, respectively~\cite{Frizyuk2019,Sautter2019}. Specifics of the radiation characteristics in each case can be explained by different parities of nonlinearly generated multipoles, so-called nonlinear multipolar interference. This represents 
an example where multipolar analysis has been proven to be a useful 
instrument for design of directional nonlinear light emission~\cite{xu2019forward}.

 A recently 
 suggested approach 
 to trap light 
 in individual subwavelength dielectric nanoresonators 
 is based on high-quality supercavity modes associated with the physics of quasi bound states in the continuum (quasi-BIC)~\cite{hsu2016bound,Rybin2017,Bogdanov2019}.  
 These states are in some sense similar to BICs in infinite periodic dielectric structures: their high finesse is due to the destructive interference of several far-field radiation channels. 
 Two interpretations of quasi-BIC formation in the open resonator, based on (i) leaky modes supported by the particle~\cite{Rybin2017} and (ii) multipoles of the electromagnetic field~\cite{Poddubny2018arXiv,Chen2019LPR}, were discussed, however no direct correspondence between these two models has been established up to now. 
 Lately, it has been proposed to utilize these high-quality 
 states to enhance the classical nonlinear process of second-harmonic generation in dielectric nanodisks~\cite{Carletti2018giant}. 
  
In this paper, we develop a comprehensive multipolar theory of  the second-harmonic generation (SHG) from high-quality quasi-BIC states in a AlGaAs cylindrical nanoresonator (nanodisk). 
We show that the formation of quasi-BIC states can be naturally understood through multipolar transformations of coupled leaky axially symmetric modes supported by the nanodisk. The strong SHG can be expected in the case of efficient excitation of the mode with the
high quality factor. To achieve efficient coupling to the mode at the pump frequency, multipolar composition of the pump source should match the multipolar structure of this mode. 
We show that azimuthally polarized beam  can be used to couple to the quasi-BIC state most efficiently because it contains magnetic multipoles with zero azimuthal numbers only 
that matches the quasi-BIC multipolar origin and maximizes the modal overlap.
Analysing spatial distribution of the induced nonlinear source with the approach 
put forward in our earlier works~\cite{Smirnova2018,Frizyuk2019}, we reveal that complex multipolar  composition of the second-harmonic radiation coincides with the multipolar composition of a particular high-quality eigenmode of the disk at double frequency. 
We obtain that the efficiency of SHG is 
strongly enhanced up to several per cent,
provided the generated frequency matches the supported resonance, for parameters of the nanoresonator  corresponding to the nearly resonant excitation of the quasi-BIC state at the fundamental frequency. 
  We perform full-wave numerical simulations of SHG 
  taking into account nonlinear effects of back-action and propose the BIC-inspired design of a hybrid metal-dielectric nanoantenna where the effect of pump depletion is further increased suggesting a promising application for the frequency downconversion.
\section{Multipolar model of quasi-BIC formation}

We consider a high-index cylindrical dielectric resonator which supports leaky modes (modes of an open resonator) that may hybridize (couple) when tuning geometric parameters. The particle is characterized by a frequency-dependent dielectric permittivity $\eps (\omega)$ and is surrounded by homogeneous host medium with $\eps_{\text{h}}=1$. The harmonic time dependence of the fields in the form ${e^{i\omega t}}$  is implied.

\begin{figure}[b!]
  \includegraphics[width=0.5\textwidth]{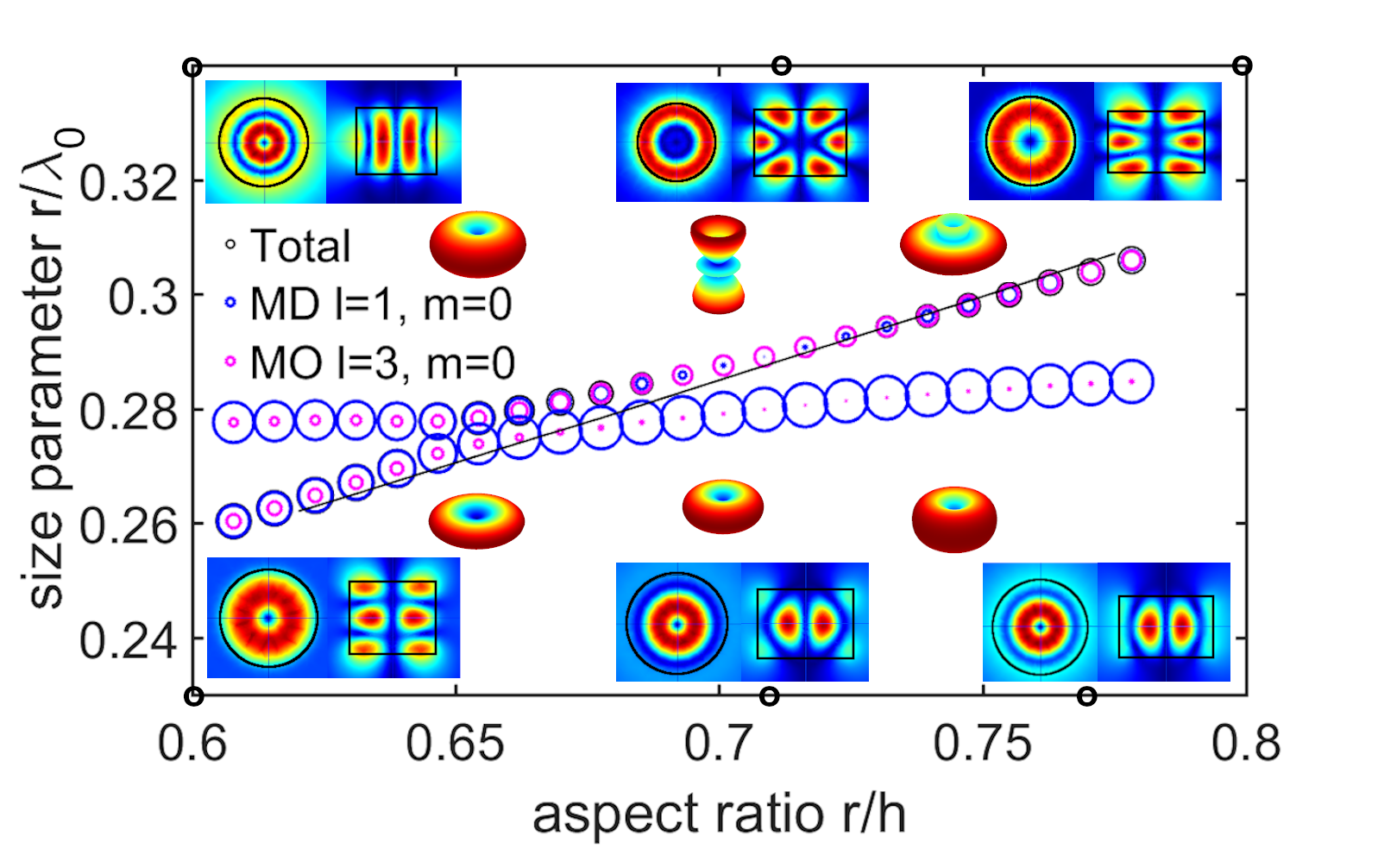}
 \caption{Dispersion of axially symmetric eigenmodes of the AlGaAs nanodisk. Colored circles depict multipolar decomposition of these modes: magnetic dipole (MD) and magnetic octupole (MO), sizes of circles correspond to their relative contributions and the radiative decay. The upper branch traces the high-quality mode and features a special point in the parameter space (${r}/{h}=0.71$,  ${r}/{\lambda_0}=0.29$), where the quality factor of this mode reaches its maximum, which corresponds to the pure MO contribution. Insets show near-field distributions of the electric field magnitude and far-field diagrams: the top row corresponds to the high-quality mode at ${r}/{h}=0.6$, ${r}/{h}=0.71$, ${r}/{h}=0.8$, the bottom row corresponds to the low-quality mode at ${r}/{h}=0.6$, ${r}/{h}=0.71$, ${r}/{h}=0.77$. The respective values of aspect ratio are marked by dots at the horizontal axis. The overlaid black line sketches a dispersion branch of the mode near the doubled frequency.
 }
 \label{fig:2}
 \end{figure}

Here we focus on rotationally symmetric TE-polarized modes in the cylindrical coordinate system $\left(E_{\phi}, H_{\rho},  H_{z}\right)$, whose electromagnetic field does not depend on azimuthal angle $\phi$. With the use of finite element method (FEM) in COMSOL Multiphysics, we perform the eigenmode analysis numerically and plot dispersion as a function of the normalized geometric parameters defined as the disk aspect ratio ${r}/{h}$ and the size parameter ${r}/{\lambda_0}$. The results of calculations for the AlGaAs nanodisk with fixed height $h=645$~nm are summarized in Fig.~\ref{fig:2}. Two dispersion curves depicted as colored dots exhibit characteristic avoided crossing in the plane of parameters, which is a signature of the strong coupling regime. It is accompanied by modification of the modes quality factors and formation of the quasi-BIC state that can be naturally understood through multipolar transformations of coupled modes as we describe in the following. 

%%%%%%%%%%%%%%%%%%%%%%%%%%%%%%
 \begin{figure}[t]
  \includegraphics[width=0.48\textwidth]{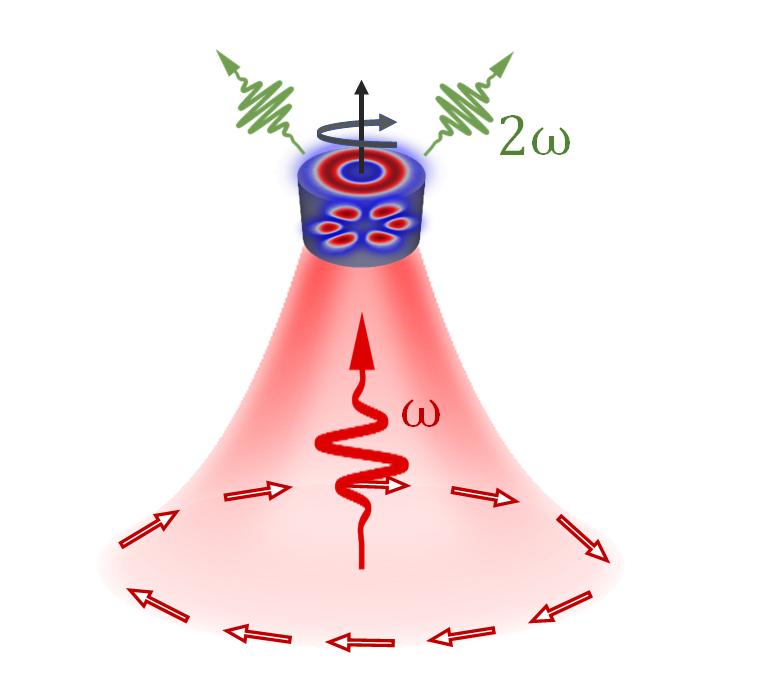}
 \caption{Schematic of the geometry. Azimuthally polarized cylindrical vector beam of frequency $\omega$ excites an axially symmetric supercavity mode in the nanodisk.
As a result of nonlinear interaction, the second-harmonic light of
$2\omega$ is generated.}
 \label{fig:1}
 \end{figure}
%%%%%%%%%%%%%%%%%%%%%%%%%%%%%%

For a nonspherical shape of the resonator the interacting modes can be in general viewed as superposition of the spherical multipoles distinguished by orbital $l$ and azimuthal $m$ indices. Multipolar analysis of the considered TE-polarized azimuthally symmetric eigenmodes suggests that a basis of parent multipoles is constituted mainly by longitudinal magnetic dipoles (MD) $l=1, m=0$ and magnetic octupole (MO) $l=3, m=0$ so that each mode has two multipolar radiation channels, dipolar and octupolar. In Fig.~\ref{fig:2}, sizes of circles illustrate their relative contributions in multipolar expansions of the modes. The occurrence of high-$Q$ supercavity mode is accompanied by increasing the order of a dominating multipole from $l = 1$ (MD) to $l = 3$ (MO) and corresponds to the decoupled magnetic octupole. At quasi-BIC condition (${r}/{h}=0.71$ and  ${r}/{\lambda_0}=0.29$) two magnetic dipoles interfere destructively in the coupling to the octupole, thus, restoring its high-quality factor. Insets show how the electric field distributions inside the disk and far-field diagrams of these modes change as we move along the dispersion curves. While far from the BIC point the radiation patterns are dipolar, near the BIC-point the pattern of the high-$Q$ mode turns to rotationally symmetric magnetic octupole with three lobes. In addition, the black line in Fig.~\ref{fig:2} shows dispersion of the mode near the doubled frequency. 
The eigenmode analysis is performed for the AlGaAs disk taking into account the material dispersion. The refractive index in the fundamental wavelength range 1500-1700~nm $n(\omega)=\sqrt{10.73}\approx 3.27$ and in the SH wavelength range 750-850~nm $n(2\omega) \approx 3.52$~\cite{Aspnes1986}.

%%%%%%%%%%%%%%%%%%%%%%%%%%%%%%
%%%%%%%%%%%%%%%%%%%%%%%%%%%%%%

\begin{figure}[h]
  \includegraphics[width=0.4\textwidth]{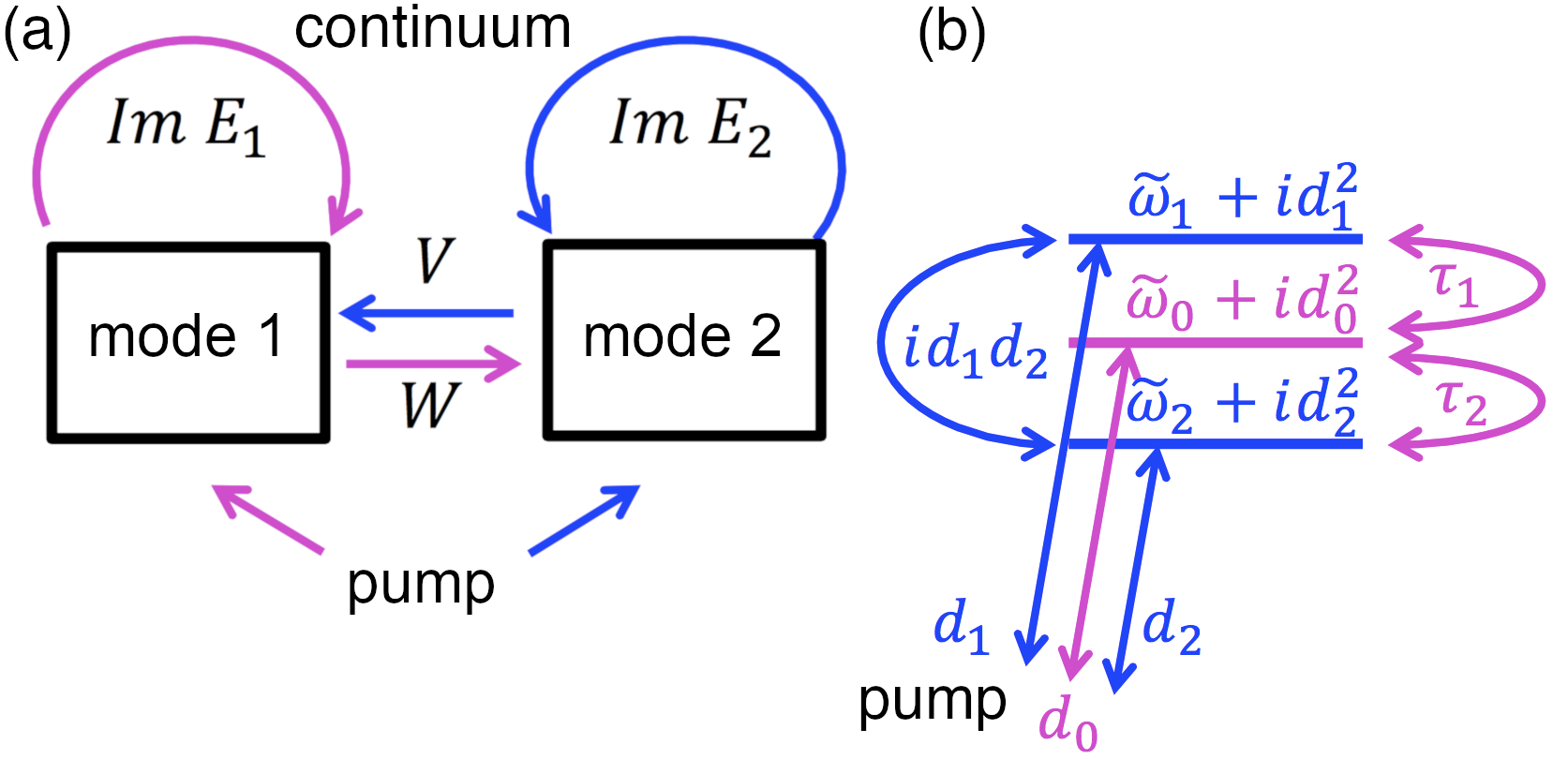}
  \caption{(a) Schematic of interaction between two coupled eigenmodes of an open dielectric resonator. Arrows illustrate individual couplings of the modes to the continuum and couplings between the modes via the continuum. (b) Schematic  of  the  three-level  multipolar model  for  the  formation of the quasi-BIC state captured by the Hamiltonian Eq.~\eqref{eq:H_3}}
  \label{fig:3}
  \end{figure}
  
We then corroborate our rigorous numerical results by an analytical model setting the correspondence of  the non-Hermitian coupled-mode theory and multipolar structure. Coupling of two leaky eigenmodes shown in Fig.~\ref{fig:2} can be described
by the Hamiltonian~\cite{Rybin2017,Bogdanov2019}
\begin{equation}
    \hat{H}_2=\begin{pmatrix}
        E_1 & V\\
        W & E_2\\
\end{pmatrix},
\end{equation} 
being a square matrix, where
$E_1$ and $E_2$ are complex frequencies of the modes $1$ and $2$, $W$ and $V$ are complex coupling parameters including interaction between the modes via free space, $W\ne V^*$. A scheme of this phenomenological two-level model is shown in Fig.~\ref{fig:3}(a). The eigenfrequencies depend on the aspect ratio  $\xi \equiv r/h$ and can be deviated from the BIC-point at $\xi^*=0.71$:  $\omega_{1,2}(\xi) = \bar{\omega} + \tilde{\omega}_{1,2}(\xi)$,  $\bar{\omega} = \omega_1(\xi^{*})= \omega_2 (\xi^{*})$. Also the frequency of incident radiation $\omega_s$ may differ from the resonance frequency $\omega_s = \bar{\omega} + \Omega$. For our system, we derive the dynamic equations for two modes with slowly varying amplitudes $B_1$ and $B_2$ excited by the external pump source in the following form:
%%%%%%%%%%%%%%%%%
\begin{equation}
\dfrac{d}{dt} \begin{pmatrix}
B_1\\
B_2
\end{pmatrix}-i\hat{H}_2\begin{pmatrix}
B_1\\
B_2
\end{pmatrix}=\begin{pmatrix}
 d_1 S_\text{d} + d_{01} S_0 \\
 d_2 S_\text{d} + d_{02} S_0  
    \end{pmatrix}e^{i\Omega t},
\end{equation}
%%%%%%%%%%%%%%%%%%%%%%%
where the matrix $\hat{H}_2$ explicitly reads
\begin{equation} 
\hat{H}_2=\begin{pmatrix}
 \tilde{\omega}_1 (\xi) + i \left[d_1^2 + d_{01}^2\right]& i\left[ d_1 d_2 + d_{01}d_{02} \right]\\
i\left[ d_1 d_2 + d_{01}d_{02} \right] &  \tilde{\omega}_2 (\xi) + i \left[d_2^2 + d_{02}^2\right]
\end{pmatrix}\:.
\end{equation}
%%%%%%%%%%%%%%%%%%%%%%%
Here $d_{1,2}$ are the effective dipole moments, and $d_{01,02}$ are effective octupole moments in the modes, $S_{\text{d}}$ and $S_{0}$ are dipolar and octupolar contributions in the incident radiation respectively. The $d$ coefficients govern  both the radiative decay  of  the  states  and  their coupling to each other and to the external field.  

Using the scattering-matrix method~\cite{Poddubny2018arXiv}, coupled-mode equations can be mapped onto the three-state model schematically illustrated in Fig.~\ref{fig:3}(b). The basis includes two magnetic-dipole states $|1\rangle$, $|2\rangle$ with $m= 0$ and the magnetic-octupole state $|0\rangle$ with $m=0$.
The evolution equations recast as
\begin{equation} \label{eq:TLM_H3}
-i\dfrac{d {\bf a}}{d t} =  \hat{H}_3 {\bf a} - i {\bf S} e^{i\Omega t},
\end{equation}
with the amplitude column-vector ${\bf a} = [a_1,a_2,a_3] $, and the external source \begin{equation}{\bf S} = \begin{pmatrix}
d_1 S_\text{d}\\
d_2 S_\text{d}\\
-e^{i\pi/4}\sqrt{\tilde{\omega}_0(\xi^*)} S_0
\end{pmatrix}.
\end{equation}
The non-Hermitian Hamiltonian of the structure in Eq.~\eqref{eq:TLM_H3} assumes the form
\begin{align}
 \hat{H}_3=\begin{pmatrix}
\tilde{\omega}_{1}(\xi)+i d_{1}^{2}& i d_{1}d_{2}&\tau_{1}\\
i d_{1}d_{2}&\tilde{\omega}_{2}(\xi)+i d_{2}^{2}&\tau_{2}\\
\tau_{1}&\tau_{2}&\tilde{\omega}_{0}(\xi)+i d_{0}^{2}&\\
\end{pmatrix},\:\label{eq:H_3}
\end{align}
where the condition $|\tilde{\omega}_0(\xi^*)|>d^2_{1,2} >d_0^2$ is satisfied,
and parameters are expressed as follows $\tau_1 =d_{01} e^{-i\pi/4}\sqrt{\tilde{\omega}_0(\xi^*)}$, $\tau_2 = d_{02}  e^{-i\pi/4} \sqrt{\tilde{\omega}_0(\xi^*)}$, $d_0^2 = {(d_{01} d_2 - d_{02}d_1)^2}/\left({d^2_1+d^2_2}\right)$.
 The $\tau$ coefficients describe the  hybridization  of the  dipole  and  octupole  modes. Ohmic  losses  are  neglected. 

The excitation of the quasi-BIC mode depends on how efficiently the incident source couples to the BIC state, that is on their spatial overlap. The external source can be divided into octupolar $S_0$ and dipolar $S_\text{d}$ components and their fractions determine the resultant response. We consider two types of pump radiation which carry $H_z$ component of the electromagnetic field and can excite the quasi-BIC mode: a linearly polarized plane wave ${\bf E}=E_0e^{ikx}{\bf \hat{y}}$ (PW) and azimuthally polarized cylindrical vector beam (AP)~\cite{Youngworth2000} (see Supporting Information).
We perform multipolar decomposition of the incident radiation in terms of vector spherical harmonics (see Supporting Information) and 
after numerical integration we obtain the following relations of the magnetic octupolar and magnetic dipolar contributions:
$\left|\frac{A_\text{M}(3,0)}{A_\text{M}(1,0)}\right|^2=0.875$ for the PW excitation, and $\left|\frac{A_\text{M}(3,0)}{A_\text{M}(1,0)}\right|^2=4.8545$ for AP excitation.
Remarkably, AP vector beam can be used to couple to the BIC mode more efficiently than linearly polarized plane wave because AP beam  can  be  decomposed solely to  magnetic  multipoles with $m$=0 that matches the multipolar composition of the modes, while the plane wave consists of electric and magnetic multipoles with different azimuthal indices $m$. Furthermore, the relative contribution of MO is significantly higher in the AP vector beam than in the plane wave, so that the azimuthally polarized cylindrical vector beam is favorable for excitation of the high-quality resonance in the dielectric nanodisk. Thus, hereafter we focus on the study of SHG in a AlGaAs nanodisk excited by AP cylindrical vector beam in the vicinity of the BIC-point. 

%%%%%%%%%%%%%%%%%%%%%%%%%%%%%
\section{Nonlinear response driven by quasi-BIC state}

For noncentrosymmetric materials the induced nonlinear polarization is defined by the second-order polarizability tensor $\hat{\chi}^{(2)}$:
\begin{equation}
{\bf P}^{(2\omega)}=\eps_0\hat{\chi}^{(2)}\bf{E}^{(\omega)}\bf{E}^{(\omega)},
\label{eq:Pnl}
\end{equation}
where $\eps_0$ is the vacuum dielectric constant, $\bf{E}$ is the electric field inside the particle. 
For AlGaAs in the principal axis system of the crystal, the tensor of the second-order nonlinear susceptibility contains only off-diagonal elements $\chi^{(2)}_{ijk}=\chi^{(2)}$ = 100 pm/V being nonzero if any of two indices $i, j, k$ do not coincide. 
Thus, the induced nonlinear polarization at the second-harmonic frequency takes the  form
\begin{equation}
    \begin{pmatrix}
P_x^{(2\omega)}\\
P_y^{(2\omega)}\\
P_z^{(2\omega)} 
\end{pmatrix}=2\eps_0\chi^{(2)}\begin{pmatrix}
E^{(\omega)}_yE^{(\omega)}_z\\
E^{(\omega)}_xE^{(\omega)}_z\\
E^{(\omega)}_xE^{(\omega)}_y
\end{pmatrix}
\label{eq:P2omega}
\end{equation}
We assume that the main axes of crystalline lattice are oriented along the Cartesian coordinate system: $[100]\parallel \hat{x},[010]\parallel \hat{y},[001]\parallel \hat{z}$. Thereby, in the case of excitation of azimuthally polarized modes inside the nanodisk the nonlinear current at double frequency has the form
\begin{equation}
{\bf j}^{(2\omega)}= 2i\omega {\bf P} =2i\omega\eps_0\chi^{(2)} E^{(\omega)2}_{\varphi} \sin 2\varphi  \bm{\hat{z}}. 
\end{equation}
Here $ E^{(\omega)}_{\varphi} $ is the electric field distribution inside the particle, which can be approximated 
by the following expression
\begin{equation}
E^{(\omega)}_{\varphi} \approx   B_1 \alpha_1  J_1(k_0 \sqrt{\eps} \rho) + B_2 \alpha_2  J_1\left(k_0\rho \sqrt{\eps-\frac{\pi}{k_0 h}}  \right) \cos \dfrac{\pi z}{2 h}, 
\end{equation}
where $J_1$ is the first order Bessel function of the first kind, $k_0={\omega}/{c}$, $\rho=\sqrt{x^2+y^2}$ is a radial distance, $\alpha_{1,2}$ are coefficients of the eigenmodes contributions.
Next, using approximation for the field profile inside the particle we analyse the multipolar composition of the SH radiation generated by the nonlinear source following
the procedure described in Ref.~\cite{Smirnova2018}. We deduce that the orbital numbers $l$ of generated multipoles are even for electric and odd for magnetic multipoles, while the azimuthal number takes the values $m = \pm 2$ only. We find four dominant generated multipoles: $ a^{(2\omega)}_{\text{M}} (2, \pm 2)$, $ a^{(2\omega)}_{\text{M}} (4, \pm 2)$, $a^{(2\omega)}_{\text{E}} (3,\pm 2)$, $a^{(2\omega)}_{\text{E}} (5,\pm 2)$. 
Our analytical calculations show that at $k_0 r \sim 1$ electric multipolar coefficients $a^{(2\omega)}_{\text{E}} (l,\pm 2)$ rapidly decrease for $l \ge 7$, while magnetic amplitudes 
$a^{(2\omega)}_{\text{M}} (l,\pm 2)$ decay fat for $l \ge 6$. For example, in the vicinity of the BIC-point we obtain the following relations in the case of AP beam excitation:
\begin{equation}
\left| \dfrac{a^{(2\omega)}_{\text{E}} (7,\pm 2)} {a^{(2\omega)}_{\text{E}} (5,\pm 2)} \right| \sim 0.02, \quad 
\left|\dfrac{a^{(2\omega)}_{\text{M}} (6,\pm 2)} {a^{(2\omega)}_{\text{M}} (4,\pm 2)}  \right| \sim 0.08.
\end{equation}

%%%%%%%%%%%%%%%%%%%%%%%%%%%%%%
\begin{figure}[h]
  \includegraphics[width=0.5\textwidth]{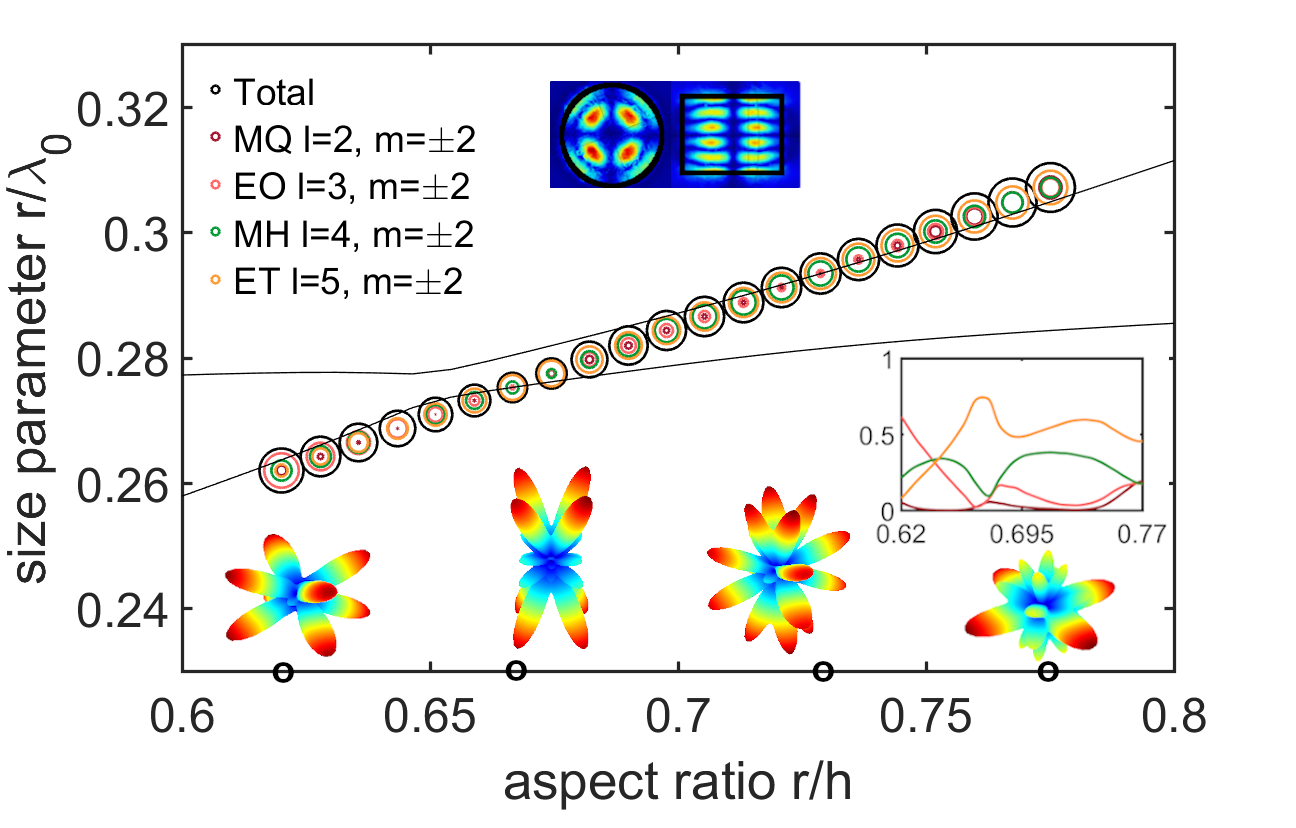}
 \caption{Dispersion and multipolar decomposition eigenmode of the AlGaAs disk at double frequency. Colored circles show multipolar decomposition of these modes, size of circles correspond to  the radiative decay and relative multipoles contribution shown in the top left inset additionally. Black solid lines are dispersion curves of eigenmodes in Fig.\ref{fig:2}.  Insets show near-field distributions of the electric field at $\frac{r}{h}=0.729$ and far-field diagrams at $\frac{r}{h}=0.62$, $\frac{r}{h}=0.667$, $\frac{r}{h}=0.729$ and $\frac{r}{h}=0.775$.
 E/M--electric/magnetic multipoles,  Q--quadrupole, O--octupole, H--hexadecapole, T--triacontadipole. }
 \label{fig:4}
 \end{figure}
 %%%%%%%%%%%%%%%%%%%%%%%%%%%%

In the case when nanoantenna supports a quasi-BIC resonance near the pump frequency and another eigenmode near the SH frequency we expect a resonant increase in the up-conversion.
In addition to two eigenmodes at the fundamental frequency, Fig.~\ref{fig:4} features the dispersion of the high-quality mode at double frequency lying in the same parameters range (SH mode). The $Q$-factor of this mode is large ($Q \approx 200-600$) (see SI). Certainly, there are several eigenmodes at double frequency in the considered parameter ranges. However, here we specifically focus on the eigenmode that can be excited by our nonlinear source ${\bf j}^{(2\omega)}= 2 i \omega  {\bf P}$, % \dfrac{d{\bf P}^{(2\omega)}}{dt}
namely, only the eigenmode with azimuthal indices $m=\pm 2$. This mode is composed of the multipoles that the induced nonlinear current at the quasi-BIC conditions generates. The multipolar decomposition of SH mode is
visualized by circles of different colors. The inset additionally shows relative contributions of multipoles in the line plot. SH mode has a more complex multipolar content than FF modes, and more sophisticated near-field distribution, which modifies slightly in the considered parameter range. The radiation pattern is generally multi-lobed and noticeably changes depending on the multipolar composition. 
The most efficient SH generation can be reached at the resonant conditions, close to crossing of BIC mode and SH mode. 

%%%%%%%%%%%%%%%%%%%%%%%%%%%%%%%%%%%%%%%%%%%%%%
%%%%%%%%%%%%
\begin{figure}[h]
  \includegraphics[width=0.5\textwidth]{{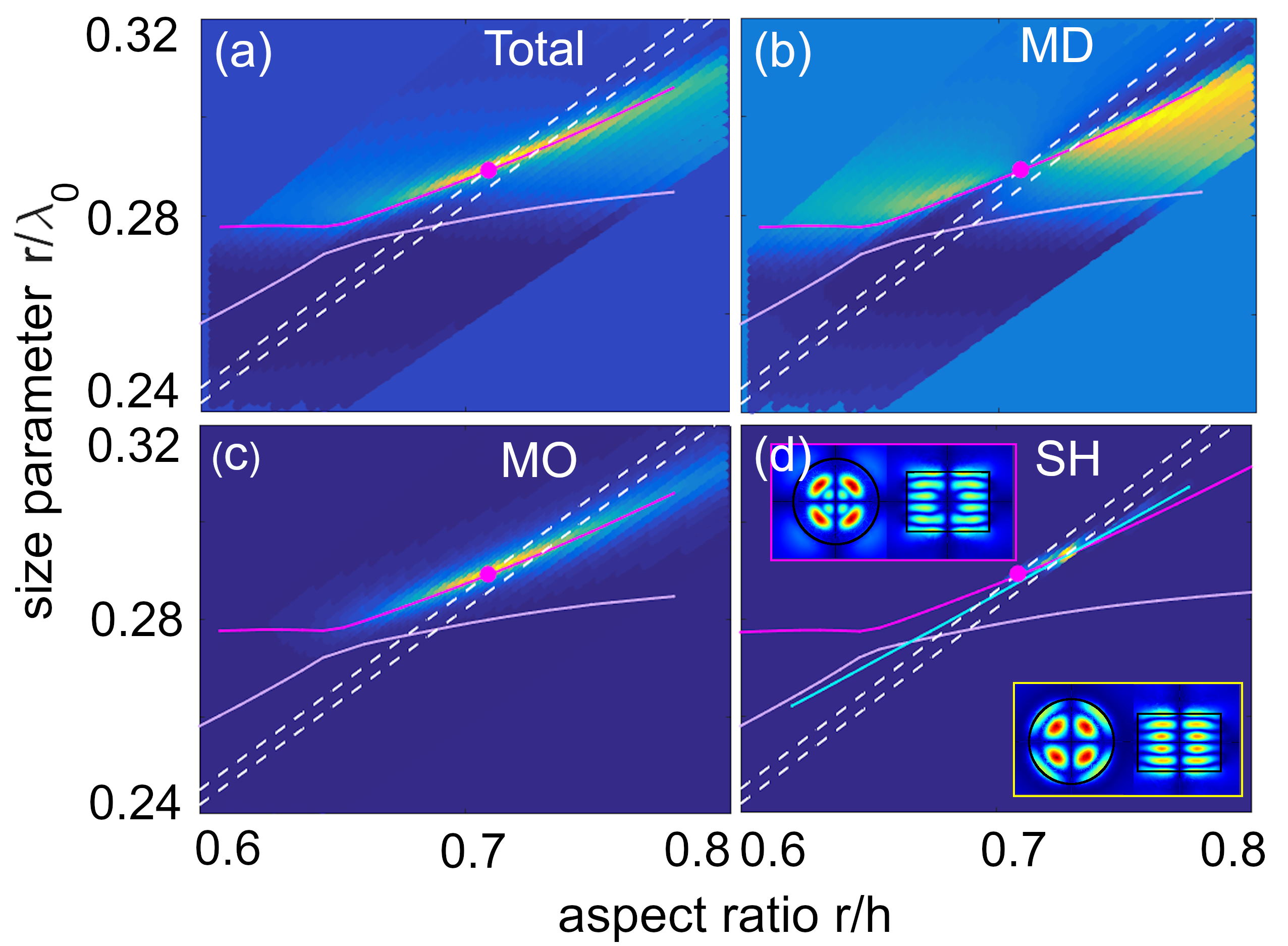}}
 \caption{Maps of linear scattering of AP beam on the disk (a) and SH generated power (d). Panels (b) and (c) corresponds to magnetic dipole and octupole contribution to the linear scattering. Solid pink lines correspond to dispersion curves of FF eigenmodes, solid blue line corresponds to dispersion curve of SH mode. Pink dot is a BIC state. Dashed lines are two slices of $\lambda_0$=1580~nm and $\lambda_0$=1600~nm. Insets in the panel (d) show electric field distribution at the BIC point (in the pink frame) and at point of intersections of dispersion curves (at maximum SH power) (in the yellow frame)}
 \label{fig:5}
 \end{figure}
%%%%%%%%%%%%%%%%%%%%%%%%

As a pump, we choose AP beam in order to maximize
overlaps of the modes with the incident field. We numerically calculate dependencies of multipolar composition of linear scattering and SH generated field on the disk radius in the case of AP beam excitation (for comparison with the case of PW excitation see Supplementary Information). The multipolar contents of both FF and SH fields appear essentially the same as we expected from the theory that justifies its validity. 
Figure~\ref{fig:5} shows two-dimensional maps of linear scattering of AP beam and SH power. We recover maps of linear scattering, MD and MO contributions in linear scattering on the nanodisk radius by calculating slices at fixed pump wavelengths. We observe strong excitation of magnetic octupole (MO) in the proximity of the BIC point so that the high-$Q$ mode branch can be clearly traced in the map of linear scattering. Next, we model SHG and obtain a sharp enhancement peak in SH intensity. The resonant area in the parameter space is visualized with a high contrast in Fig.~\ref{fig:5}(d). These our results fully explain the recent experimental observations reported in Refs.~\cite{Koshelev2019CLEO,Koshelev2019arXiv}.

%%%%%%%%%%%%%%%%%%%
\begin{figure}[h]
  \includegraphics[width=0.5\textwidth]{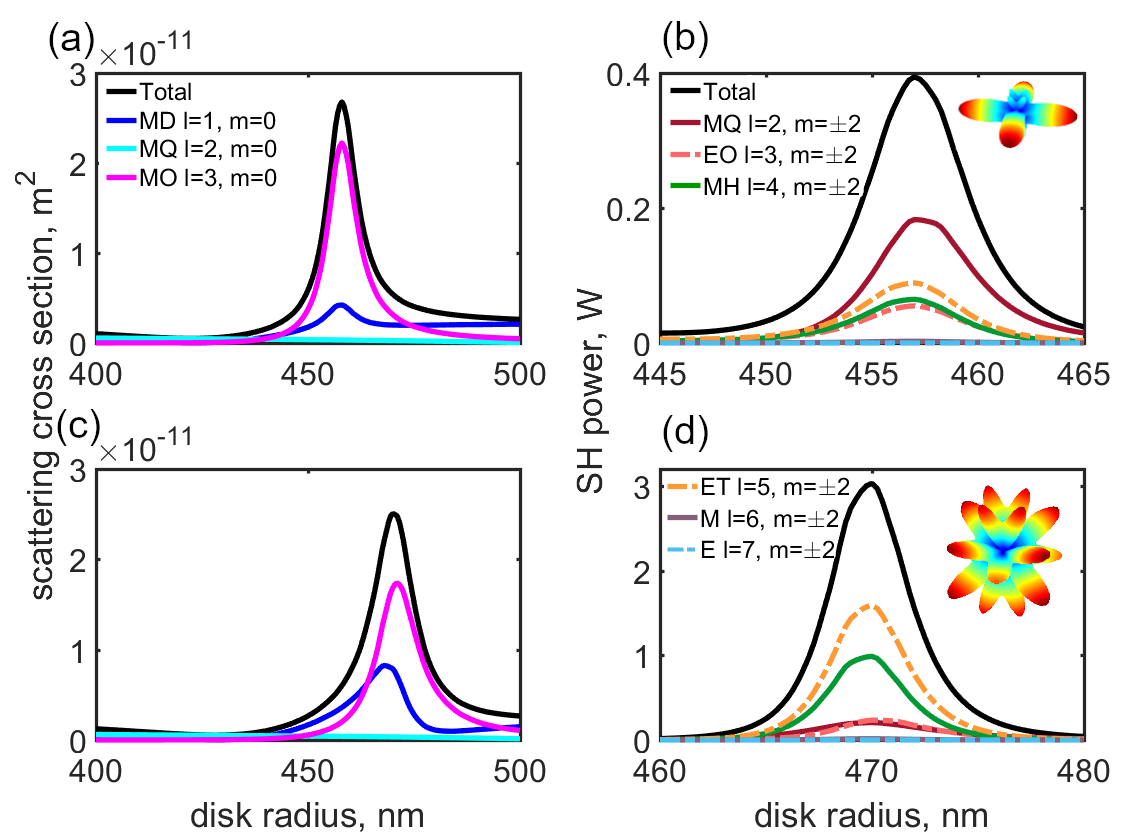}
 \caption{Multipolar decomposition of linear scattering (a,c) and second-harmonic generation (b,d) spectra for two slices in Fig.\ref{fig:5}. Top and bottom panels correspond to pump wavelengths $\lambda_0=1580$~nm and $\lambda_0=1600$~nm, respectively. Colored lines show different multipole contributions to the linear scattered and the second-harmonic fields. Insets show far-field diagrams of SH radiation at BIC-point ($\lambda_0=1580$~nm, $r=457$~nm, $h=645$~nm) and at the point of dispersion curves intersection ($\lambda_0=1600$~nm, $r=470$~nm, $h=645$~nm). E/M--electric/magnetic multipoles, D--dipole, Q--quadrupole, O--octupole, H--hexadecapole, T--triacontadipole.
 }
 \label{fig:6}
 \end{figure}
%%%%%%%%%%%%%%%%%%%%%%%%%%%%%%%

Figure~\ref{fig:6} shows two slices of 2D color maps in Fig.~\ref{fig:5} for two close wavelengths: the slice at $ \lambda_0=1580$~nm contains the BIC point and the slice at $\lambda_0=1600$~nm contains the intersection of SH and FF dispersion curves. If we consider the fixed pump wavelength $\lambda_0=1580$~nm, 
we see that MO dominates in the linear scattering and the maxima of the scattering cross-section and SH power for this slice correspond to the BIC point ($r=457$~nm, $h=645$~nm). We obtain that in this case magnetic quadrupole (MQ) $a^{(2\omega)}_{\text{M}} (2,\pm 2)$ dominates in the  emission, and the SH radiation pattern is mainly quadrupolar. For the wavelength $\lambda_0$=1600~nm corresponding to the intersection of the FF high-$Q$ mode and SH mode dispersion branches, the SHG enhancement is even higher. In this case major multipoles are $a^{(2\omega)}_{\text{M}} (4,\pm 2)$, $a^{(2\omega)}_{\text{E}} (5,\pm 2)$,
that matches the multipolar content of the SH mode.
%%%%%%%%%%%%%%%%%%%%%%%

Because the distribution of the AP beam intensity is inhomogeneous, here we define the scattering efficiency as the ratio of the scattering power at fundamental frequency $P_{\text{FF}}$ to the energy flux through the geometrical cross section of the particle in the waist plane $P_{\text{AP}}$:
    \begin{eqnarray}
          &\tilde{\sigma} =\dfrac{P_{\text{FF}}}{P_{\text{AP}}}, \\
      & P_{\text{AP}}=\int{\bf \overline{S}}{\bf n}dS=-\int \dfrac{E_{\phi}H_{\rho}^{*}+E_{\phi}^{*}H_{\rho}}{4}dS.
  \end{eqnarray}
Here, ${\bf \overline{S}}=\frac{1}{2} Re\left( \bf {E} \times \bf{H^{*}}\right)$ is a time-averaged Poynting vector, ${\bf n}\parallel \hat{\bf z}$ is a normal vector.
The scattering cross section $\sigma_{\text{sca}}=\tilde{\sigma}S$ is much larger than the geometric cross section $S$ ($S=\pi r^2$ for AP and $S=2rh$ for PW) of the disk near the BIC-point especially for AP beam (${\sigma}/{\pi r^2}\approx 40$) because of the high quality-factor of MO resonance. In this case it would be incorrect to define conversion efficiency as the ratio of generated power $P_{\text{SH}}$ and $P_{\text{AP}}$ because these values become comparable near the BIC point. But SH generated power is much smaller than scattered power at fundamental frequency because of large quality factor of excited BIC mode.
Thus, to take into account the strong interaction of the incident radiation with the resonance mode we define the second-harmonic conversion efficiency as the ratio of the total SH radiated power $P_{\text{SH}}$ to the radiated power at the fundamental frequency $P_{\text{FF}}$:
 \begin{equation}
      \rho=\frac{P_{\text{SH}}}{P_{\text{FF}}}. 
  \end{equation}
  We obtain that the absolute maximum of SHG corresponds to intersection point and ideally, it may reach about 1 \% conversion.
  Also we calculate conversion efficiency that doesn't depend on the incident power \begin{equation}
      \tilde{\rho}=\frac{P_{\text{SH}}}{P_{\text{FF}}^2}.
      \label{eq:effcy2}
  \end{equation}
 The simple coupled-mode considerations suggest scaling the SHG efficiency (\ref{eq:effcy2}) depending on detunings from the resonances at the fundamental and SH frequencies: 
\begin{equation}
   \tilde{\rho} \sim  Q_{\text{FF}}^2 Q_{\text{SH}} \left( \frac{\gamma_{\text{FF}}^2}{(\omega -\omega_{\text{FF}})^2+\gamma_{\text{FF}}^2}\right)^2 \frac{\gamma_{\text{SH}}^2}{(2\omega -\omega_{\text{SH}})^2+\gamma_{\text{SH}}^2}.
   \label{eq:rho}
\end{equation}
Here, $Q_{\text{FF}}$ and $Q_{\text{SH}}$ are quality factors of resonances, $\omega_{\text{FF}}$ and $\omega_{\text{SH}}$ are resonance frequencies, $\gamma_{\text{FF}}$ and $\gamma_{\text{SH}}$ are damping constants. 
The highest yield of SHG can be attained in the double-resonant case.
%%%%%%%%%%%%%%%%%%%%%%%%%%%%%%%%%%

\begin{figure}[h]
  \includegraphics[width=0.5\textwidth]{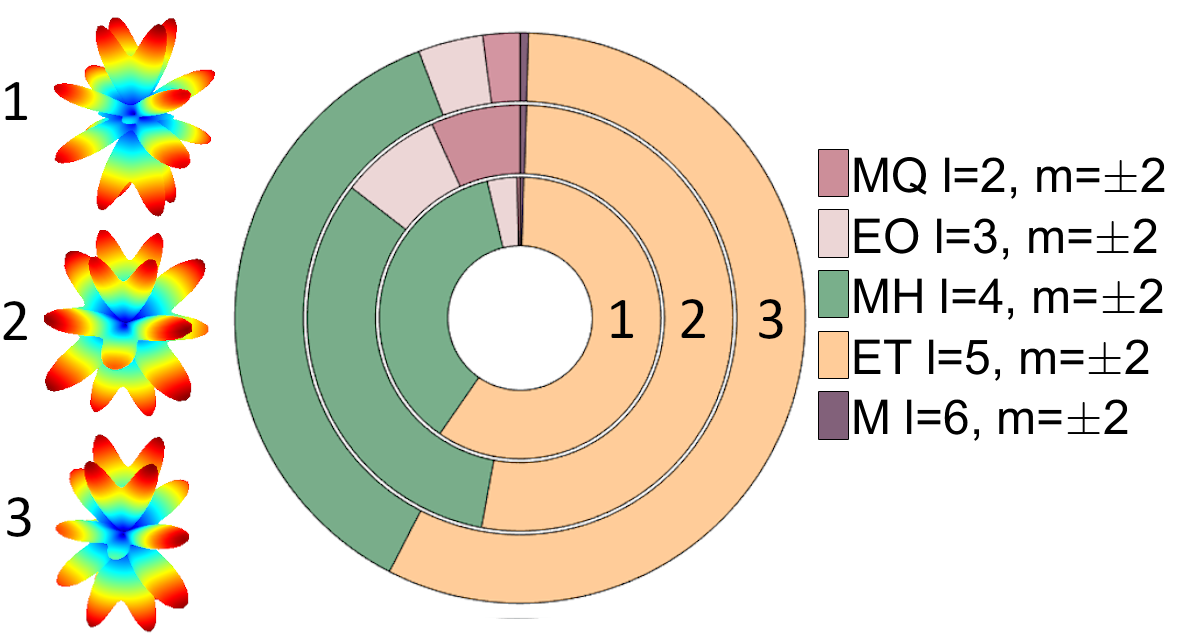}
 \caption{Comparison of far field diagrams and multipolar composition of  second-harmonic-generated electromagnetic field and the disk's eigenmode at double frequency for parameters of the maximum SHG efficiency  ($\lambda_0$=1600~nm, $r$=470~nm, $h$=645~nm): 1 - SH mode, 2 - AP beam excitation, 3 - PW excitation. E/M--electric/magnetic multipoles, Q--quadrupole, O--octupole, H--hexadecapole, T--triacontadipole.}
 \label{fig:7}
 \end{figure}

We compare near-field profiles, far-field diagrams and multipolar compositions of SH mode and SH radiation near the peak SHG for two types of the incident radiation. We obtain that at the parameters of the maximum SHG, the multipolar composition of the SH generated field, electric field distribution and radiation pattern are almost the same as those of the SH eigenmode, see Figs.~\ref{fig:4},~\ref{fig:5},~\ref{fig:7}. 
This explicitly shows that even though the generated SH radiation is composed of several different multipoles, the only one SH mode characterized by $m=\pm2$ is predominantly generated. 

%%%%%%%%%%%%%%%%%%%%%%%%%%%%%%%%%%%%%%%%%%%%%%%%
 \section{Coupled model of SHG with pump depletion} \label{sec:PumpDeplModel}
 
In our numerical modeling we take into account the nonlinear effects of back-action of the second-harmonic radiation on the fundamental frequency radiation. In the case of highly-efficient double-resonant excitation the electric field amplitudes inside the particle at the fundamental and doubled frequencies may become comparable, and the nonlinear correction of the polarization ${\bf P}_{nl}^{(\omega)}$ at the fundamental frequency can no longer be neglected. Thus, to simulate the back-action nonlinear effect, we numerically solve the Helmholtz equation for electric fields with  nonlinear sources
at the fundamental and SH frequencies simultaneously:
\begin{equation}
    -\Delta {\bf E}+\mu_0 \dfrac{\partial^2 {\bf D
     }}{\partial t^2}=-\mu_0\dfrac{\partial^2}{\partial t^2}\left({\bf P}_{\text{nl}}^{(\omega)}+{\bf P}_{\text{nl}}^{(2\omega)}\right)\:.
\end{equation}
The expression for nonlinear polarisation at fundamental frequency ${\bf P}_{\text{nl}}^{(\omega)}$ in the principle crystalline axis system has the following form:
\begin{equation}
    \begin{pmatrix}
P_x^{(\omega)}\\
P_y^{(\omega)}\\
P_z^{(\omega)} 
\end{pmatrix}=2\eps_0\chi^{(2)}\begin{pmatrix}
E^{(2\omega)}_yE^{(\omega)*}_z+E^{(2\omega)}_zE^{(\omega)*}_y\\
E^{(2\omega)}_xE^{(\omega)*}_z+E^{(2\omega)}_zE^{(\omega)*}_x\\
E^{(2\omega)}_xE^{(\omega)*}_y+E^{(2\omega)}_yE^{(\omega)*}_x
\end{pmatrix}\:.
\end{equation}
The nonlinear source at the second-harmonic frequency is given by Eq.~\eqref{eq:P2omega}. 

At the conditions of the resonant excitation of the magnetic-octupole quasi-BIC mode and high-quality mode at the second-harmonic frequency, the electric field can be represented as a superposition 
\begin{equation}
    {\bf E}=\epsilon_1 (t) {\bm {\epsilon}}_1{\left({\bm r}\right)}e^{i\omega_1 t}+ \epsilon_2 (t) {\bm {\epsilon}}_2{\left( \bm r \right)}e^{i\omega_2 t}+c.c.\:,
\end{equation}
where $\epsilon_{1,2} (t)$ are the time-dependent amplitudes, $|\epsilon_{1,2}|^2$ are the energies of the modes at the frequencies $\omega_{1,2}$,
$\Delta \omega = \omega_2 - 2\omega_1$, $|\Delta \omega| \ll \omega_{1,2}$ is a detuning from the synchronism.  
Assuming the amplitudes slowly varying in time,  with the characteristic time scale $\tau=\left|\dfrac{1}{\epsilon_{1,2}}\dfrac{d\epsilon_{1,2}}{dt} \right|^{-1} \gg
\dfrac{2\pi}{\omega_{1,2}}$, one can write truncated equations
\begin{equation}
    \begin{split}
        \dfrac{\partial {{\epsilon} }_1}{\partial t} + \gamma_1 \epsilon_1  =- i { \bar{\chi}}{{\epsilon} }_2{\ {\epsilon} }_1^* e^{i\Delta \omega t} + \sqrt{\gamma_1} S_0 e^{i\Omega t}, \\
        \dfrac{\partial {{\epsilon} }_2}{\partial t} + \gamma_2 \epsilon_2 = - i  {\bar{\chi}}^{*}{{\epsilon} }^2_1 e^{-i\Delta \omega t},
    \end{split}
\end{equation}
where the coefficient 
\begin{equation}
\bar{\chi} \sim 2 \omega \chi^{(2)} \int \left( \epsilon_{1y} \epsilon_{1z} \epsilon^{*}_{2x} + \epsilon_{1x} \epsilon_{1z} \epsilon^{*}_{2y} + \epsilon_{1x} \epsilon_{1y} \epsilon^{*}_{2z}    \right) d^3{\bm r} \:
\end{equation}
is proportional to the overlap integral, $\gamma_{1,2}$ are decay rates of the quasi-BIC and SH modes.

We calculate scattering and SHG conversion efficiencies taking into account coupling of excited SH and FF modes and compare them with the numbers  obtained in the undepleted pump approximation. The results for two types of incident radiation for parameters corresponding to the maximum SHG efficiency ($\lambda_0$=1600~nm, $r$=470~nm, $h$=645~nm) are summarized in Table~\ref{tab:effcy}. 
%%%%%%%%%%%%%%%%%%%%%
\begin{table}[h]
    \centering
    \begin{tabular}{c|c|c|c}
  & $\tilde {\sigma}$ & $\rho$  & ${\tilde \rho},W^{-1}$\\
    AP undepleted pump & 34.14  & 0.02 & $1.33\cdot 10^{-4}$\\
    AP back action & 32.93 & 0.0182 & $1.25\cdot 10^{-4}$\\
     PW undepleted pump & 3.97 &0.0031&$5.49\cdot 10^{-5}$\\
     PW back action & 3.95 &0.0031&$5.44\cdot 10^{-5}$\\
    \end{tabular}
    \caption{Scattering and second harmonic conversion efficiencies}
    \label{tab:effcy}
\end{table}
%%%%%%%%%%%%%%%%%%%%%%%%%
Values of the pump intensity and energy flux through the geometric cross section of the disk used in our calculations are given in Table~\ref{tab:pump}.
%%%%%%%%%%%%%%%%%%%%%
\begin{table}[h]
    \centering
    \begin{tabular}{c|c|c|c|c}
  & $I_{\text{min}},~\frac{\text{GW}}{\text{cm}^2}$ & $I_{\text{max}},~\frac{\text{GW}}{\text{cm}^2}$  &$S\cdot 10^{13},~\text{m}^2$& $P$,~\text{W}\\
    AP & 0.89 & 2.16 & 6.94& 4.404\\
    PW & 2.37& 2.37& 6.06 &14.36\\
    \end{tabular}
    \caption{Pump characteristics}
    \label{tab:pump}
\end{table}
%%%%%%%%%%%%%%%%%%%%%%%%%

We obtain that in the case of AP excitation nonlinear effects of back action become noticeable due to efficient excitation of the high-quality modes. Thus, 
coupling of electromagnetic fields at fundamental and second harmonic frequencies should be taken into consideration when the conversion efficiency reaches 
1\%. 
For the PW excitation, the maximum SHG is also reached for the parameters corresponding to the intersection of the eigenmodes dispersion curves, however, the total generated power turns to be lower than in the case of AP beam pump. As we noticed above, it can be explained using multipolar analysis of the incident radiation. The multipolar composition of the incident plane wave is more complicated than multipolar composition of AP beam, the spectrum of linear scattering contains several multipoles with different values of azimuthal index $m$. As a result, the electric field distribution at the BIC point is distorted and appears dependent on azimuthal angle $\phi$. We still observe a resonance at the BIC point in the linear scattering spectrum but the quality factor of this resonance is lower, and the corresponding scattering efficiency is an order of magnitude lower than in the case of AP excitation. Also, as we have shown, the magnetic octupolar relative contribution with $m=0$ is smaller in the plane wave, thereby the excitation of the quasi-BIC state is less efficient. 

In the next section, we consider the BIC-inspired nanoantenna design where the second-harmonic radiation significantly influences the linear scattering. Such nanoresonators can be particularly promising for realization of the efficient frequency downconversion at the nanoscale. 

\section{BIC-inspired nonlinear enhancement through high-quality MD resonance 
via AP beam illumination}

One type of BICs which has been widely studied is the symmetry-protected BICs existing at $\Gamma$ point of a periodic system. 
In periodic system, when the coupling of a certain resonance to the radiation modes are forbidden by symmetry mismatch, a symmetry-protected BIC is formed~\cite{hsu2016bound}. Such symmetry-protected BICs cannot be excited directly under normal plane wave incidence, while they can be excited externally by breaking the in-plane $C_2$ symmetry of the system, for example, by introducing a defect in the nanodisks to open a radiation channel and transform the ideal-BIC to quasi-BIC with finite $Q$ factor. In particular, by properly designing a nanodisk metasurface, it can support resonant longitudinal magnetic dipole resonance at $\Gamma$ point at the frequency below the diffraction limit. Such MD mode does not couple to the free-space radiation channel due to the symmetry mismatch, forming a BIC. For single nanoresonators, the out-of-plane Mie-type magnetic modes can be excited efficiently via azimuthally-polarized beam~\cite{das2015beam}, whereas for periodic systems this is not possible because the input field for each unit cell has to be a structured field distribution. However, inspired by this symmetry-protected BIC mechanism, here we suggest a novel approach to design high-quality mode in isolated Mie nanoresonators by lifting the restriction on the periodic boundary conditions.

%=================================================
\begin{figure}[t!]
	\centering
	\includegraphics[width = 1.0\columnwidth]{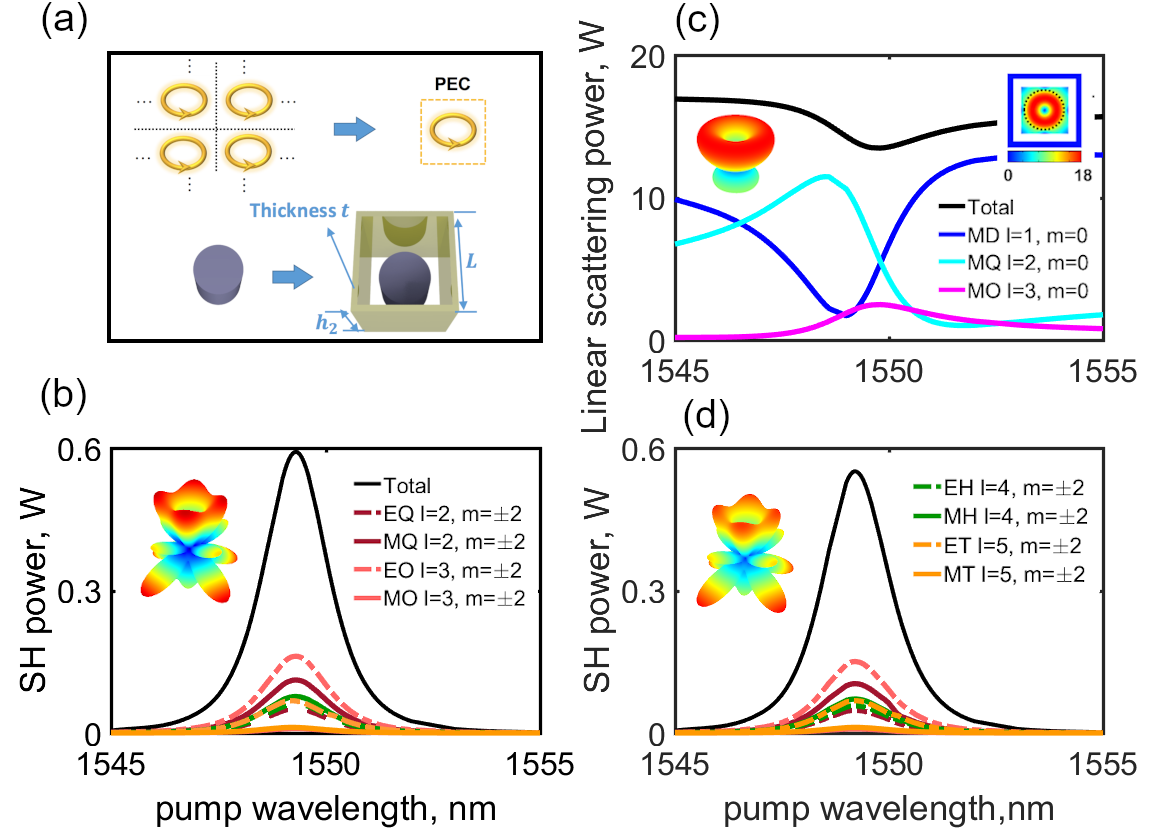}
	\caption{(a) Top: Schematic of the equivalent transformation between circular polarization currents excited in a periodic lattice of MD resonators and currents excited in a single MD resonator surrounded by PEC structures. Bottom: The transformation from a 2D periodic array of MD resonators to an isolated nanostructure composed of a MD resonator surrounded by Au nanobars. (b) The calculated linear scattering and multipolar decomposition under AP pump illumination for the AlGaAs-disk-with-Gold nanostructure. (c) SHG emission power and relative multipolar contributions as a function of the pump wavelength calculated using (c) the undepleted pump model, and (d) the coupled nonlinear model with back-action. E/M--electric/magnetic multipoles, D--dipole, Q--quadrupole, O--octupole, H--hexadecapole, T--triacontadipole.}
	\label{fig:SPBIC}
\end{figure}
%=================================================

A schematic illustration of the proposed principle is shown in Fig.~\ref{fig:SPBIC}(a). 
As was described in Ref.~\cite{xu2018boosting}, 
when an electric or polarization current is placed near a PEC surface, the excited free electron oscillations in the adjacent PEC affect the near-field and far-field properties of such composite system. The effect can be reproduced with an oppositely oriented image of the electric current by using the image dipole model. Using this analogy, a periodic system of MD resonators can be transformed to as a single MD resonator surrounded by a PEC-like box. As an example, we employ a dielectric disk surrounded by Au nanobars, as shown in Fig.~\ref{fig:SPBIC}(b). Remarkably, by shrinking the periodic system into a single nanoresonator, the quasi-BIC MD mode can be directly and efficiently excited under structured beam illumination that further facilitates
nonlinear frequency conversion.

We then take the AlGaAs nanodisk supporting a magnetic dipole resonance near 1550~nm as and study the nonlinear performance of our proposed design. The geometric parameters of our structure are determined to be the following: the radius and height of AlGaAs nanodisk are $r_0=$237.5~nm, $h_0=$400~nm; the length, height and thickness of the gold structure are $L=$575~nm, $h_2=2h_0$ and $t=$100~nm. We suppose an azimuthally-polarized beam, with the maximum intensity $\mathrm{I_{\text{max}} = 2.16}$~$\mathrm{GW/cm^2}$, focused by an objective with the numerical aperture NA$=0.9$ is illuminating our sample. Figure~\ref{fig:SPBIC}(b) shows the calculated linear scattering.
A narrow  dip appears around 1550~nm corresponding to excitation of the high-$Q$ quasi-symmetry-protected-BIC MD mode. 
The quality-factor of this supercavity mode is estimated to be $Q=554$ which is much larger than the typical MD resonance supported by an individual nanoresonator ($\sim$ 9)~\cite{marino2019spontaneous}. Its longitudinal MD nature enables optimal overlapping with a AP pump.

We further examine the SHG process from the designed AlGaAs/Gold hybrid nanoresonator. In order to show the importance of our advanced simulation method introduced in Sec.~\ref{sec:PumpDeplModel}, we compare the calculated results based on the undepleted pump model and the coupled back-action model. 

In the undepleted pump model, we follow two steps to simulate the nonlinear response. First, we calculate the linear response at the fundamental wavelength, and obtain the nonlinear polarization induced inside the structure based on the formula (\ref{eq:P2omega}). We then employ the nonlinear polarization as a source to simulate the electromagnetic response at the harmonic wavelength. In this two-step procedure, we disregard the influence of the harmonic waves on the pump field. Figures~\ref{fig:SPBIC}(c) and (d) show the calculated SHG power and the corresponding nonlinear multipolar contributions for different pump wavelengths using the undepleted pump model and the coupled nonlinear model with back-action, respectively.
As can be seen, the SHG emission can be significantly enhanced when exciting the designed high-$Q$ resonance at the pump frequency. 
We observe a clear difference in the nonlinear emission power when using the coupled back-action model as compared to the undepleted pump model: at the pump wavelength of 1549~nm, the total SHG power is around 0.593~W and 0.546~W for these two cases, respectively. This indicates that neglecting  coupling from SH field to the pump field in the undepleted pump model  results in more than 8\% error in the nonlinear simulation. 
Under AP pump illumination, the SHG signal can be boosted significantly in our proposed configuration leading to nearly 2,000-fold enhancement of the SHG emission power as compared to the case of SHG driven by the conventional MD resonance in a single free-standing nanodisk. 

%%%%%%%%%%%%%%%%%%%%%%%%%
\section{Conclusion}
%%%%%%%%%%%%%%%%%%%%%%%%%
  We have explicated the multipolar nature of quasi-BIC states in subwavelength dielectric resonators supporting high-quality supercavity modes called quasi-bound states in the continuum and discussed  their applications for  nonlinear nanophotonics. Using our multipolar model, we have analyzed optimal conditions for the efficient excitation of quasi-BIC states in high-index dielectric nanodisks under structured light illumination. In particular, we have explained the multifold increase of the second-harmonic conversion efficiency for the case of azimuthally polarized cylindrical vector beam illumination compared to the linearly polarized plane wave excitation. Implementing numerically the coupled nonlinear model, we have clarified values of SHG efficiencies and discussed pump depletion in isolated high-quality nanoresonators.

 \section*{Acknowledgements}
This work is supported by the Russian Foundation for Basic Research (Grant Nos. 18-02-00381 and 19-02-00261). I.V. acknowledges partial support from the Foundation for the Advancement of Theoretical Physics and Mathematics BASIS. The work of A.E.M. is supported by a UNSW Scientia Fellowship.

\end{document}